\documentclass[useAMS,usenatbib,times,mathptm]{mn2e}

\usepackage{umlaut}
\usepackage{color}
\usepackage{graphicx}
\usepackage{graphics}
\usepackage{rotating}
\usepackage{amsmath}        
\usepackage{amssymb}        

\newcommand{\degree}{$^{\circ}$}

\newcommand{\perday}{\,d$^{-1}$}

\title[Deceleration of the jets in GRS\,1915+105?]{Evidence for
deceleration in the radio jets of GRS\,1915+105?}

\author[J.C.A.~Miller-Jones et al.]
 {J.C.A.~Miller-Jones,$^1$\thanks{email: jmiller@science.uva.nl}
 M.P.~Rupen,$^2$ R.P.~Fender,$^{3,1}$ A.~Rushton,$^4$ G.G.~Pooley,$^5$\and and
 R.E.~Spencer.$^4$\\
$^1$Astronomical Institute `Anton Pannekoek', University of Amsterdam,
 Kruislaan  403, 1098 SJ, Amsterdam, The Netherlands\\
$^2$NRAO, Array Operations Center, 1003 Lopezville Road, Socorro, NM
 87801, U.S.A.\\
$^3$School of Physics and Astronomy, University of Southampton,
Highfield, Southampton, SO17 1BJ\\
$^4$Jodrell Bank Observatory, The University of Manchester,
 Macclesfield, Cheshire, SK11 9DL\\
$^5$Astrophysics, Cavendish Laboratory, J.~J.~Thomson Avenue, Cambridge CB3 0HE
}
\begin{document}

\date{Accepted xxxxxxx. Received xxxxxxx; in original form xxxxxxx}

\pagerange{\pageref{firstpage}--\pageref{lastpage}} \pubyear{2006}

\maketitle

\label{firstpage}

\begin{abstract}
There is currently a clear discrepancy in the proper motions measured
on different angular scales in the approaching radio jets of the black
hole X-ray binary GRS\,1915+105.  Lower velocities were measured with
the Very Large Array (VLA) prior to 1996 than were subsequently found
from higher-resolution observations made with the Very Long Baseline
Array and the Multi-Element Radio Linked Interferometer Network.  We
initiated an observing campaign to use all three arrays to attempt to
track the motion of the jet knots from the 2006 February outburst of
the source, giving us unprecedented simultaneous coverage of all
angular scales, from milliarcsecond scales out to arcsecond scales.
The derived proper motion, which was dominated by the VLA
measurements, was found to be 17.0\,mas\perday, demonstrating that
there has been no significant permanent change in the properties of
the jets since 1994.  We find no conclusive evidence for deceleration
of the jet knots, unless this occurs within 70\,mas of the core.  We
discuss possible causes for the varying proper motions recorded in the
literature.

\end{abstract}

\begin{keywords}
ISM: jets and outflows -- stars: winds, outflows -- radio
continuum:stars -- stars:individual:GRS1915+105 -- stars:variables --
X-rays: binaries
\end{keywords}

\section{Introduction}
\label{sec:intro}
Since they were first observed by \citet{Mir94}, the radio jets in the
black hole X-ray binary GRS\,1915+105 have been the focus of intensive
study.  The ejecta have been observed at milliarcsecond scales with
the Very Long Baseline Array (VLBA) \citep{Dha00} and the
Multi-Element Radio Linked Interferometer Network (MERLIN)
\citep{Fen99,Mil05}, and at arcsecond scales with the Very Large Array
(VLA) \citep{Mir94,Rod99}.  \citet{Rod98} and \citet{Kai04} have also
searched on arcminute scales for potential impact sites of the jets on
the interstellar medium, although at present the evidence is
inconclusive.

\begin{figure}
\begin{center}
\includegraphics[width=\columnwidth,angle=0,clip=]{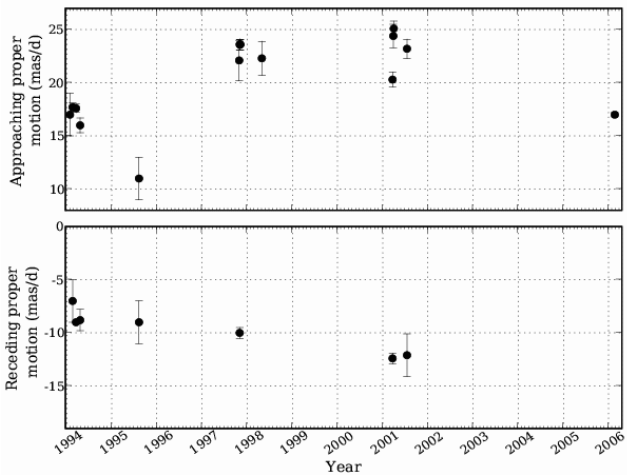}
\caption{Measured proper motions in GRS\,1915+105, as a function of
  time.  The August 1995 point was made from only a single pair of
  individual condensations, so was considered unreliable \citep{Rod99}.}
\label{fig:epochs}
\end{center}
\end{figure}

The proper motions initially measured by \citeauthor{Mir94} with the
VLA were $\mu_{\rm app}=17.6\pm0.4$\,mas\,d$^{-1}$ for the approaching
(southeastern) jet and $\mu_{\rm rec} = 9.0\pm0.1$\,mas\,d$^{-1}$ for
the receding (northwestern) counterjet.  \citet{Rod99} used the VLA to
measure proper motions for other ejection events, and found fairly
consistent values, with no evidence for deceleration of the jets,
which were in all cases found to move ballistically outwards from the
core.  However, MERLIN observations of a 1997 ejection event
\citep{Fen99} measured a significantly higher set of proper motions,
$\mu_{\rm app} = 23.6\pm0.5$\,mas\,d$^{-1}$ and $\mu_{\rm rec} =
10.0\pm0.5$\,mas\,d$^{-1}$, albeit on smaller angular scales.  Again,
the jet knots were observed to move ballistically.  Such high proper
motions were also confirmed by VLBA observations \citep{Dha00} and
further MERLIN observations \citep{Mil05}.  In no case was there ever
any evidence for deceleration.  Fig.~\ref{fig:epochs} shows all the
published proper motions for the jets in this system as a function of
time, and highlights the stark discrepancy between observations
prior to 1997 made with the VLA and those made later with
higher-resolution arrays.  The actual data are listed in Table
\ref{tab:propermotions}.

\begin{table*}
\begin{minipage}{0.95\textwidth}
\renewcommand{\thefootnote}{\thempfootnote}
\caption{Published proper motions for the jets in GRS\,1915+105.}
\begin{center}
\begin{tabular}{lccll}
\hline \hline
Outburst date & $\mu_{\rm app}$ (mas\,\,d$^{-1}$) & $\mu_{\rm rec}$
(mas\,\,d$^{-1}$) & Instrument & Reference \\
\hline
29 Jan 1994 & $17\pm2$ &  & VLA & \citet{Rod99} \\
19 Feb 1994 & $17.7\pm0.4$ & $7\pm2$ & VLA & \citet{Rod99} \\
19 Mar 1994 & $17.6\pm0.4$ & $9.0\pm0.1$ & VLA & \citet{Mir94} \\
21 Apr 1994 & $16.0\pm0.7$ & $8.8\pm1.0$ & VLA & \citet{Rod99} \\
10 Aug 1995 & $11\pm2$\footnote{Since this proper motion was based on only a
  single pair of individual condensations, it was considered unreliable.} & $9\pm2$\footnotemark[\value{mpfootnote}] & VLA & \citet{Rod99} \\
28 Oct 1997 & $22.1\pm1.9$\footnote{Converted from the proper motions per hour
  quoted by \citet{Dha00}.} &  & VLBA & \citet{Dha00} \\
29 Apr 1998 & $22.3\pm1.6$\footnotemark[\value{mpfootnote}] &  & VLBA & \citet{Dha00} \\
01 Nov 1997 & $23.6\pm0.5$ & $10.0\pm0.5$ & MERLIN & \citet{Fen99} \\
04 Nov 1997 & $23.6\pm0.5$ &  & MERLIN & \citet{Fen99} \\
06 Nov 1997 & $23.6\pm0.5$ &  & MERLIN & \citet{Fen99} \\
21 Mar 2001 & $20.3\pm0.7$\footnote{It was found that the positional
  uncertainties in \citet{Mil05} had been overestimated, so the fits
  to the data were redone using the revised error bars to give the
  proper motions quoted here, with associated uncertainties.} &
$12.4\pm0.5$\footnotemark[\value{mpfootnote}] & MERLIN & \citet{Mil05} \\
27 Mar 2001 & $24.4\pm1.1$\footnotemark[\value{mpfootnote}] &  & MERLIN & \citet{Mil05} \\
31 Mar 2001 & $25.1\pm0.7$\footnotemark &  & MERLIN & \citet{Mil05} \\
16 Jul 2001 & $23.2\pm0.9$\footnotemark[\value{mpfootnote}] & $12.1\pm2.0$\footnotemark[\value{mpfootnote}] & MERLIN & \citet{Mil05} \\
\hline \hline
\end{tabular}
\label{tab:propermotions}
\end{center}
\end{minipage}
\end{table*}

Several possibilities have been put forward to explain this apparent
discrepancy in the motion of the jet knots between milliarcsecond
angular scales, as measured with MERLIN and the VLBA, and those seen
on arcsecond scales with the VLA.  \citet{Fen99} found that the
discrepancy could not be explained away by a simple change in the
angle to the line of sight of an otherwise physically identical jet
(i.e.\ ruling out precession).  They suggested two possible
explanations; either the jet velocities were intrinsically different,
or resolution effects were at work between the two arrays.  This
latter would lead to a blending of components, and would explain the
proper motion discrepancy if there was a sequence of ejecta which
decreased sequentially more rapidly in flux density with increasing
distance from the core (such that each ejected knot decayed more
slowly than its predecessors).  This effect was originally highlighted
by \citet{Hje95} for the case of GRO J\,1655--40.  Observing the same
outburst sequence with both the VLBA and VLA, they found a similar
reduction in the fitted proper motion from 54 to 40\,mas\,d$^{-1}$ on
moving from VLBA to VLA angular scales.  They suggested that a
combination of the flux density fading with time and spatial averaging
of underlying structures led to the discrepancy.

A further possibility is that the knots could be decelerating as they
moved outwards (which would imply that they could not be highly
relativistic after the deceleration, since the Lorentz factor $\Gamma$
must be of order 2 to see discernible changes in the proper motion),
although the ballistic motions of the observed knots would tend to
argue against this.  Alternatively, as pointed out by \citet{Mil05},
there could have been an intrinsic change in the system between the
time when the VLA observations were taken (all between 1994 and 1995)
and the time when the higher-resolution observations were made
(between 1997 and 2001).  Since the source was not detected until 1992
\citep{Cas92}, despite the existing X-ray satellites prior to that
time having the sensitivity to detect it, the system might only have
emerged from quiescence at that time.  Indeed, \citet{Tru06} suggested
that the system has been in a continuous outburst phase since 1992,
and may soon switch off again, moving back into quiescence.  In this
case, the powerful jets might since 1992 have evacuated a cavity
through which the ejecta could propagate, such that they could coast
for longer during later outbursts, giving rise to the higher proper
motions measured with the VLBA and MERLIN.  As suggested by
\citet{Hei02}, this would explain the location of the system within a
region of very low ISM density, although in that case the jet would
have needed to have been active for a timescale of order $10^3$ years,
rather than only since 1992.

To try to discern the cause of this proper motion discrepancy, we
proposed to track the motion of the ejecta in GRS\,1915+105 over all
available angular scales, using the VLBA, MERLIN and also the VLA.  In
Section \ref{sec:obs}, we detail our observations, in Section
\ref{sec:discussion} we discuss our results, and in Section
\ref{sec:conclusions} we draw our conclusions.

\section{Observations and data reduction}
\label{sec:obs}
\begin{figure*}
\begin{center}
\includegraphics[height=0.85\textheight,angle=0,clip=]{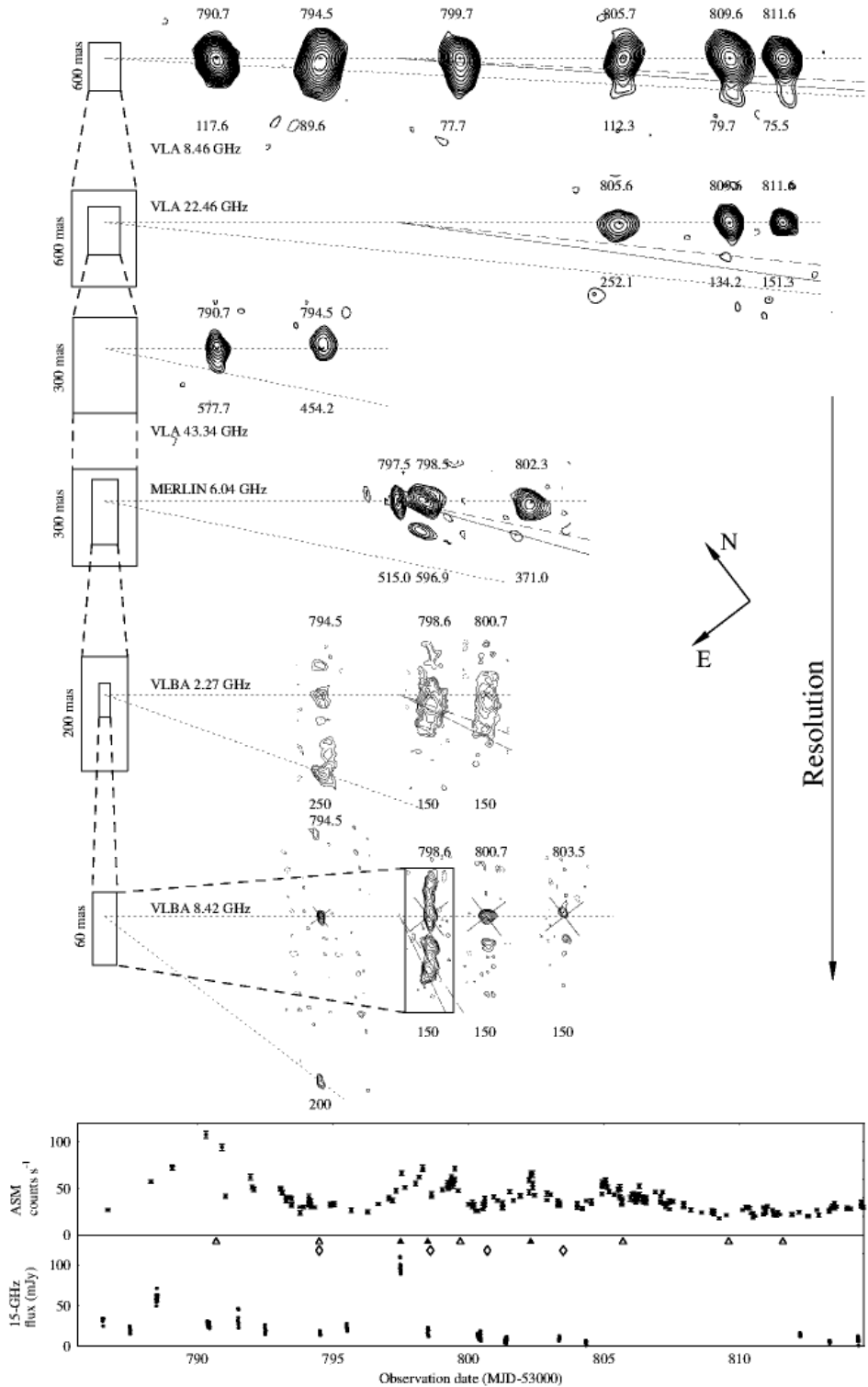}
\caption{Radio images from our monitoring campaign.  Flux densities
  from the 15-GHz Ryle monitoring programme and the 1.5--12\,keV X-ray
  count rates measured by the {\it RXTE} ASM are shown in the two
  bottom plots.  Open triangles at the top of the Ryle plot mark the
  times of the VLA observations, diamonds the VLBA observations, and
  filled triangles the MERLIN observations.  Resolution gets better on
  moving down the page, showing, in order, VLA 8.46\,GHz images (top),
  VLA 22.46\,GHz images, VLA 43.34\,GHz images, MERLIN 6.03\,GHz
  images, VLBA 2.27\,GHz images, and finally VLBA 8.42\,GHz images
  (bottom; note the different scale for the first epoch).  The dates
  of the observations (${\rm MJD}-53000$) are shown above each radio
  image, and the image rms (in $\mu$Jy\,bm$^{-1}$) below.  Contours
  are at levels of $\pm(\sqrt{2})^n$ times the rms level, where
  $n=3,4,5,...$.  The predicted position of the VLBA core (V.~Dhawan,
  private communication) is marked by a cross.  All images have been
  rotated anticlockwise by $37.5^{\circ}$.  Dotted lines show the core
  position and the fitted proper motion of 17.0\,mas\,d$^{-1}$ from
  the initial ejection event, and the dashed and dot-dashed lines show
  nominal proper motions of 23.6 and 17.0\,mas\perday\ respectively,
  for an ejection event taking place at the time of the detected
  MERLIN flare.}
\label{fig:montage}
\end{center}
\end{figure*}

We triggered our Target of Opportunity proposal \citep{Mil06} after
being alerted to a radio flare in the system by the RATAN-600
monitoring programme (S.~Trushkin, private communication).  The {\it
Rossi X-ray Timing Explorer (RXTE)} All Sky Monitor (ASM) data showed
that this coincided with a significant X-ray flare, giving us
confidence that an ejection event was underway.  We obtained four
epochs of VLBA observations, three of MERLIN observations, and
quasi-weekly monitoring with the VLA until the end of its
A-configuration trimester.  A montage of all the observations made
during the initial period when all three arrays were taking data is
shown in Fig.~\ref{fig:montage}.

\subsection{VLBA observations}
\label{sec:vlba}

The VLBA observations were made in parallel at two different
frequencies, 8.42 and 2.27\,GHz.  Observations were taken using dual
polarisation, 2-bit sampling, with sixteen 500-kHz channels giving a
total bandwidth of 8\,MHz at each frequency.

GRS\,1915+105 is scatter-broadened, with a scattering size of
$1.9\pm0.1$\,mas at 8.4\,GHz, scaling as $\lambda^2$ \citep{Dha00}.
This corresponds to a scattering size of 25.6\,mas for the longer
wavelength observations at 2.3\,GHz.  We were therefore unable to make
use of the full resolution of the array, and were forced to discard
the longest baselines (most affected by the scattering) before
imaging.  We imaged with a robust weighting parameter biased toward
natural weighting, applying a Gaussian taper in the {\it uv}-plane to
downweight the contributions of the longest remaining baselines.  

The integrated flux density of the source did not appear to vary by
more than 10--20 per cent during any of the four observing runs, so
the images were made with the full timerange available at each epoch
(4\,h for the first epoch, and 6\,h for the other three) to maximise
the {\it uv}-coverage.  With an assumed proper motion of
24\,mas\perday\ (as seen in the previous high-resolution
observations), components would move approximately one beamwidth over
the course of the observation at 8.4\,GHz and half a beamwidth at
2.3\,GHz, leading to some smearing of the components in the images.
This effect has been investigated by previous authors
\citep[e.g.][]{Mio01}, and has been shown not to significantly affect
the conclusions that can be drawn from such images.

The VLBA measurements are summarised in Table
\ref{tab:vlba_measurements}, and the images can be seen in
Fig.~\ref{fig:montage}.  Only in the first of the four epochs was the
original southeastern component seen which corresponded to the knot
monitored with the VLA (Section \ref{sec:vla}).  Its separation of
$135.9\pm0.5$\,mas at position angle $142\fdg 3\pm0\fdg 2$ agrees well
with that seen at the VLA (see Table \ref{tab:vla_measurements}) on
the same day (MJD\,53794).  Both the core and the ejecta appear
slightly elongated, and in both the 8.4 and 2.3\,GHz images there
appears to be a hint of a receding northwestern component.

\begin{table*}
\caption{VLBA observations of GRS\,1915+105.}
\begin{center}
\begin{tabular}{cccrrr}
\hline \hline
Observation date & Frequency & Component & Angular Separation &
PA & Flux density\\
(MJD) & (GHz) & & (mas) & (degs) & (mJy) \\
\hline
$53794.541\pm0.081$ & 8.4 & Core & 0 & 0 & $8.97\pm0.56$\\
$53794.541\pm0.081$ & 8.4 & SE & $135.9\pm0.5$ & $142.3\pm0.2$ &
$4.44\pm0.68$\\
$53794.541\pm0.081$ & 8.4 & NW & $68.2\pm0.8$ & $-34.0\pm0.6$ &
$1.13\pm0.20$\\
$53794.541\pm0.081$ & 2.3 & Core & 0 & 0 & $6.21\pm0.91$\\
$53794.541\pm0.081$ & 2.3 & SE1 & $16.0\pm1.3$ & $163.1\pm4.5$ &
$5.00\pm0.44$\\
$53794.541\pm0.081$ & 2.3 & SE2 & $139.6\pm0.7$ & $141.6\pm0.3$ &
$11.34\pm1.12$\\
$53794.541\pm0.081$ & 2.3 & NW & $51.5\pm2.1$ & $-41.4\pm2.3$ &
$1.94\pm0.67$\\
$53798.564\pm0.125$ & 8.4 & Core & 0 & 0 & $6.78\pm0.73$ \\
$53798.564\pm0.125$ & 8.4 & SE1 & $12.9\pm0.3$ & $142.6\pm1.4$ &
$5.61\pm0.49$ \\
$53798.564\pm0.125$ & 8.4 & SE2 & $23.2\pm0.3$ & $140.7\pm0.8$ &
$6.61\pm0.52$ \\
$53798.564\pm0.125$ & 8.4 & NW1 & $8.2\pm0.4$ & $-31.8\pm2.3$ &
$3.38\pm0.42$ \\
$53798.564\pm0.125$ & 8.4 & NW2 & $13.3\pm0.3$ & $-37.9\pm1.3$ &
$2.98\pm0.34$ \\
$53800.657\pm0.124$ & 8.4 & Core & 0 & 0 & $4.04\pm0.30$ \\
$53800.657\pm0.124$ & 8.4 & SE & $11.5\pm0.3$ & $145.3\pm1.4$ &
$2.15\pm0.39$ \\
$53800.657\pm0.124$ & 2.3 & Core & 0 & 0 & $8.53\pm0.87$ \\
$53800.657\pm0.124$ & 2.3 & SE & $48.9\pm1.5$ & $138.5\pm1.8$ &
$3.38\pm0.60$ \\
$53803.525\pm0.124$ & 8.4 & Core & 0 & 0 & $3.32\pm0.60$\\
$53803.525\pm0.124$ & 8.4 & SE & $10.9\pm0.4$ & $150.7\pm1.6$ &
$2.85\pm0.58$\\
\hline \hline
\end{tabular}
\label{tab:vlba_measurements}
\end{center}
\end{table*}

By the second epoch (MJD\,53798), the southeastern component has faded
and is no longer detectable, although there is evidence that at least
two new sets of knots have been ejected.  In the 8.4\,GHz image, we see
both the approaching and receding components, and the core component
appears to be marginally elongated, hinting at an even more recent
ejection event.  The size of the scattering disc at 2.3\,GHz prevents us
from seeing this structure as clearly at the longer wavelength,
although the source is certainly resolved.

Two days later, at MJD\,53800, these new components have also faded at
8.4\,GHz, and the image shows an unresolved core and a new, resolved
southeastern component.  At the lower frequency however, the source is
clearly resolved, and a component presumably corresponding to the
knots from the previous epoch can be seen at an angular separation of
$48.9\pm1.5$\,mas.

In the last epoch, MJD\,53803, we again see an unresolved core and a
very weak southeastern component at 8.4\,GHz.  For the known range of
proper motions of the ejecta in this system (16--24\,mas\,d$^{-1}$),
the approaching component in this image cannot be related to those
seen in the previous two images.  No source was detected at 2.3\,GHz.
For this observing run, we were missing two of the antennas in the
southwestern United States (Fort Davies and Pie Town), which removed
several of the short baselines in the array.  Coupled with the loss of
the long baselines to scattering, this rather compromised the {\it
uv}-coverage, which might explain the non-detection.

\subsection{MERLIN observations}
\label{sec:merlin}
Three epochs of MERLIN observations were taken.  The observing
frequency was 6.0353\,GHz with a bandwidth of 14 MHz.  At each epoch,
in addition to the target source GRS\,1915+105, observations were made
of the flux and polarisation angle calibrator 3C\,286, the point
source calibrator OQ\,208, and the phase reference source
B\,1919+086, 2\fdg 8 away from the target source.  The
observations were made using the five outstations (Cambridge, Defford,
Knockin, Darnhall and Tabley) and the Mark 2 antenna at Jodrell Bank.

The MERLIN d-programs were used to perform initial data editing and
amplitude calibration, and the data were then imported into the
National Radio Astronomy Observatory's (NRAO) \textsc{Astronomical
Image Processing System (aips)} software package for further data
reduction.  The MERLIN pipeline was then used to image and
self-calibrate the phase reference source, and apply the derived
corrections to the target source, GRS\,1915+105, which was subjected
to further iterations of phase-only self-calibration.

The first MERLIN epoch caught the start of a flare, seen to begin at
11:10\,UT on 2006 March 3 (MJD\,$53797.465\pm0.003$).  The variable
amplitude meant that the last few hours of observation had to be
flagged to prevent the violation of the basic assumption of synthesis
imaging that the source structure should not change during the
observation.  Similarly, the second epoch of observation caught the
decay of a further flare, dropping from 107 to 20\,mJy between 02:30
and 09:00\,UT.  Nevertheless, imaging that epoch after removing the
affected data showed both the core component and a southeastern jet
knot at an angular separation of $115.5\pm3.1$\,mas.  The final epoch
appeared to show only an unresolved core.  The images are shown in
Fig.~\ref{fig:montage}.

\subsection{VLA observations}
\label{sec:vla}

The VLA was in A-configuration, and 1-hour observations were made on a
quasi-weekly basis.  The main observing frequency was 8.46\,GHz,
although initially when the source was bright, higher frequencies were
also used in an attempt to improve the resolution.  As the ejecta
moved outwards and faded below detectability at 8.46\,GHz, lower
frequencies were used in order to take advantage of the assumed steep
spectrum ($\nu^{-0.75}$) of the jet knots to aid in their detection.

In all cases, the flux calibrator was 3C\,48, and the phase calibrator
was J\,1924+156, at an angular separation of 5\fdg 25 from the
target source.  The fast switching mode was used at frequencies above
5\,GHz in order to reduce the slew time between the secondary
calibrator and the target while cycling between the two fast enough to
account for tropospheric phase variations.  This allowed for
diffraction-limited imaging at high frequencies with the long
baselines available in A-configuration, in the case that the source
was not strong enough for self-calibration to work reliably.  Data
reduction was performed using standard procedures within
\textsc{aips}.

\begin{table*}
\begin{minipage}{0.95\textwidth}
\renewcommand{\thefootnote}{\thempfootnote}
\caption{VLA observations of GRS\,1915+105.}
\begin{center}
\begin{tabular}{crrcrr}
\hline \hline
Observation date & Frequency & Angular Separation &
PA & Core flux density & SE component flux density\\
(MJD) & (GHz) & (mas) & (degs) & (mJy) & (mJy) \\
\hline
$53761.72749\pm0.00272$ & 8.46 & $190.1\pm0.9$ & $142.5\pm0.3$ &
$18.34\pm0.28$ & $41.30\pm0.27$\\
$53761.73860\pm0.00573$ & 43.34 & $191.7\pm5.6$ & $145.8\pm1.7$ &
$23.59\pm0.93$ & $2.60\pm1.15$\\
$53765.71609\pm0.00231$ & 8.46 & $291.5\pm3.6$ & $144.9\pm0.7$ &
$5.14\pm0.25$ & $8.91\pm0.30$\\
$53765.72678\pm0.00571$ & 43.34 & & & $1.54\pm0.39$ & \\
$53790.68594\pm0.00145$ & 8.46 & $68.9\pm0.6$ & $148.0\pm0.5$ &
$26.85\pm0.12$ & $24.53\pm0.12$\\
$53790.69282\pm0.00289$ & 43.34 & $50.4\pm2.3$ & $140.8\pm2.4$ &
$18.62\pm0.65$ & $5.16\pm0.57$\\
$53794.50428\pm0.00405$ & 8.46 & $135.9\pm1.2$ & $147.6\pm0.5$ &
$12.24\pm0.09$ & $12.09\pm0.09$\\
$53794.51493\pm0.00405$ & 43.34 & & & $3.98\pm0.53$ & \\
$53799.65492\pm0.00318$ & 8.46 & $209.0\pm2.6$ & $143.8\pm0.7$ &
$8.55\pm0.09$ & $3.22\pm0.09$\\
$53805.68200\pm0.00376$ & 8.46 & $319.0\pm19.1$ & $143.8\pm3.6$ &
$11.28\pm0.15$ & $0.74\pm0.15$\\
$53805.59288\pm0.00480$ & 22.46 & & & $6.26\pm0.53$ & \\
$53809.61429\pm0.00330$ & 8.46 & $379.0\pm15.1$ & $147.6\pm2.0$ &
$13.39\pm0.08$ & $0.65\pm0.08$\\
$53809.62459\pm0.00480$ & 22.46 & $217.4\pm16.0$ & $137.1\pm4.1$ &
$7.24\pm0.25$ & $0.42\pm0.15$\\
$53811.57042\pm0.00492$ & 8.46 & $458.2\pm27.1$ & $150.7\pm2.8$ &
$12.18\pm0.14$ & $0.47\pm0.08$\\
$53811.58380\pm0.00579$ & 22.46 & & & $10.02\pm0.28$ & \\
$53817.64936\pm0.01835$ & 8.46 & $460.0\pm17.6$ & $146.5\pm1.9$ &
$6.17\pm0.06$ & $0.29\pm0.03$\\
$53824.58872\pm0.01823$ & 8.46 & $685.0\pm23.6$ & $143.4\pm1.8$ &
$8.26\pm0.06$ & $0.16\pm0.03$\\
$53831.59398\pm0.01898$ & 4.86 & $751.6\pm33.9$ & $147.4\pm2.2$ &
$3.72\pm0.09$ & $0.21\pm0.05$ \\
$53838.63336\pm0.01829$ & 8.46 & $834.9\pm48.1$ & $138.9\pm3.3$ &
$3.86\pm0.06$ & $0.09\pm0.03$\\
$53845.57801\pm0.02037$ & 1.425 & $1004.8\pm87.9$ & $152.8\pm4.2$ &
$24.20\pm0.09$ & $0.57\pm0.08$\\
$53854.62338\pm0.01991$ & 1.425 & & & $14.90\pm0.10$ & \\
$53859.62893\pm0.02766$ & 1.425 & & & $167.65\pm0.30$\footnote{In fact
  the source flux density rose from 135 to 180\,mJy during this
  observing run.} & \\
\hline
\end{tabular}
\label{tab:vla_measurements}
\end{center}
\end{minipage}
\end{table*}

\subsubsection{The flare of 2006 January}
\begin{figure}
\begin{center}
\includegraphics[width=0.5\textwidth,angle=0,clip=]{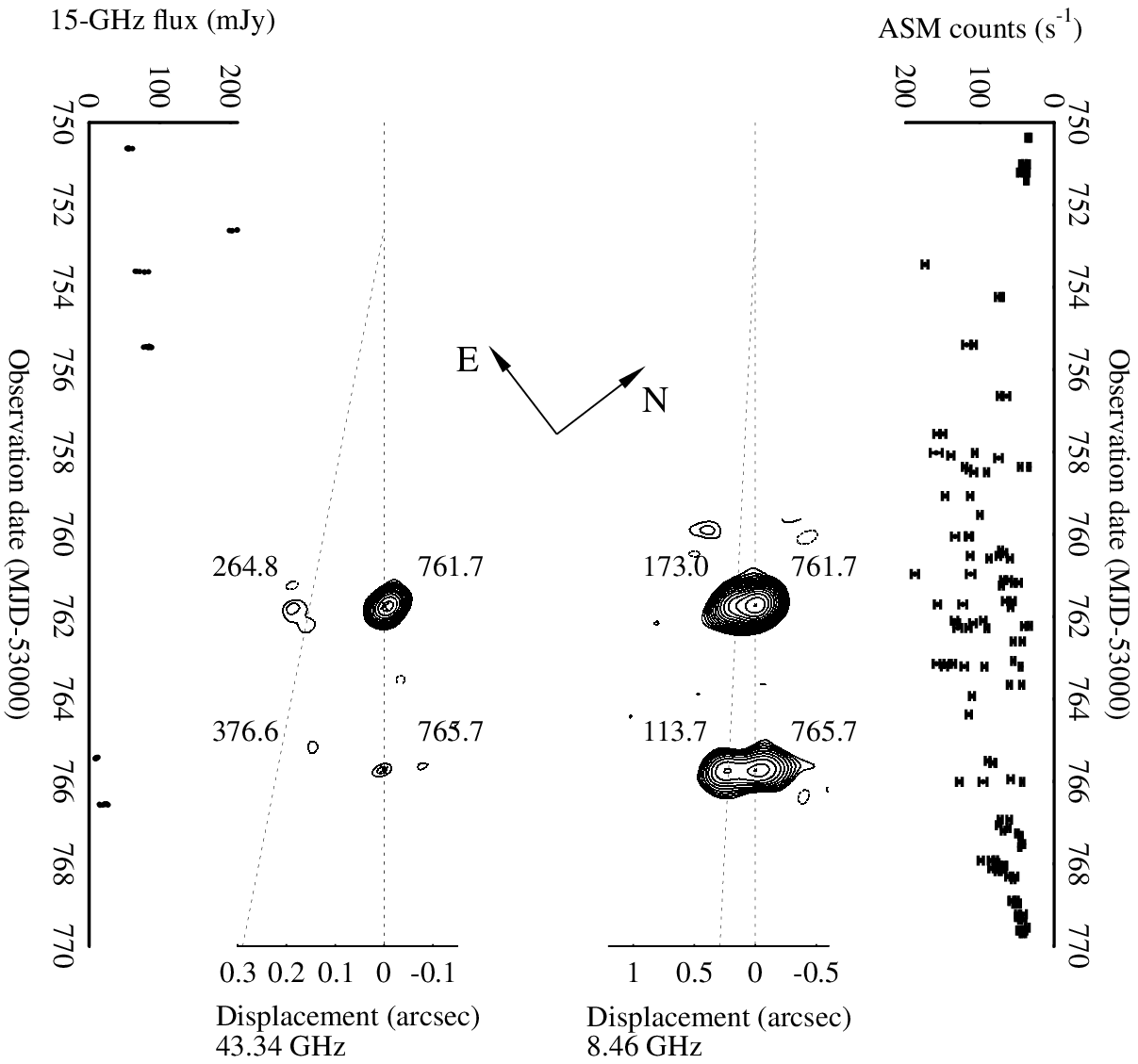}
\caption{VLA images of the 2006 January ejection event.  Flux
  densities from the 15-GHz Ryle monitoring programme are shown on the
  left-hand axis and the 1.5--12\,keV X-ray count rates measured by
  the {\it RXTE} ASM are on the right-hand axis.  The 43.34\,GHz data
  are shown in the left-hand column of images, and 8.46\,GHz images in
  the right-hand column.  The dates of the observations are shown at
  the top right of each radio image, and the image rms (in
  $\mu$Jy\,bm$^{-1}$) at the top left.  Contours are at levels of
  $\pm(\sqrt{2})^n$ times the rms level, where $n=3,4,5,...$.  The
  predicted position of the VLBA core (V.~Dhawan, private
  communication) is marked by a cross.  All images have been rotated
  clockwise by $52.5^{\circ}$.  Dotted lines show the core position
  and a nominal proper motion of 17.0\,mas\,d$^{-1}$ for a knot
  ejected at the time of the Ryle flare.}
\label{fig:jan_montage}
\end{center}
\end{figure}

The 15-GHz Ryle Telescope monitoring programme \citep{Poo97} detected
a flare from GRS\,1915+105 which peaked at 211\,mJy on MJD\,53752.6
(2006 January 17).  But since the 15-GHz monitoring was not
continuous, the start of the flare could have been as early as
MJD\,53750.6.  The source was resolved by the VLA into two components
on both MJD\,53761 and MJD\,53765.  The proper motion between these
two epochs was $25.4\pm0.9$\,mas\,d$^{-1}$.  From the fitted positions
of the two components in the maps prior to self-calibration, there is
evidence that both the northwestern and the southeastern components
moved between the two epochs, the former along position angle
$-36.7\pm3.4$\degree and the latter along $146.5\pm2.8$\degree.  This
suggests that the proper motion we see is the sum of the proper
motions of the approaching and receding components, so that of the
approaching component alone is likely to be somewhat lower.  These
images are shown separately in Fig.~\ref{fig:jan_montage}.

\subsubsection{The flare of 2006 February}
\begin{figure}
\begin{center}
\includegraphics[height=0.82\textheight,angle=0,clip=]{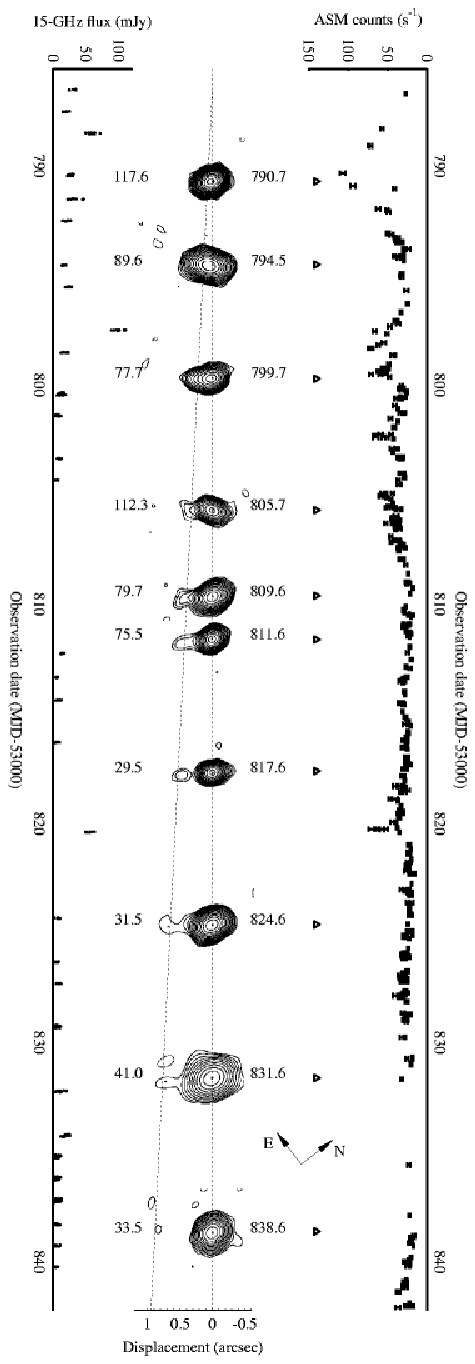}
\caption{VLA images of the 2006 February ejection event (all at
  8.46\,GHz except for the 4.86\,GHz image of MJD\,53831.6).  Flux
  densities from the 15-GHz Ryle monitoring programme are shown on the
  left-hand axis and the 1.5--12\,keV X-ray count rates from the {\it
    RXTE} ASM are on the right-hand axis.  Triangles in the X-ray plot
  mark the exact observation dates.  The dates of the observations are
  shown to the right of each image, and the image rms (in
  $\mu$Jy\,bm$^{-1}$) to the left.  Contours are at levels of
  $\pm(\sqrt{2})^n$ times the rms level, where $n=3,4,5,...$.  The
  predicted position of the VLBA core (V.~Dhawan, private
  communication) is marked.  All images have been rotated clockwise by
  $52.5^{\circ}$.  Dotted lines show the core position and the fitted
  proper motion of 17.0\,mas\,d$^{-1}$.}
\label{fig:vla_montage}
\end{center}
\end{figure}

The main radio flare on which we triggered our multi-resolution
observing campaign appeared to peak on MJD\,53788.5 (2006 February
22).  We tracked one single ejected component for 59 days with the VLA
as it moved southeast, out to an angular separation of over 1\,arcsec.
The full sequence of images is shown in Fig.~\ref{fig:vla_montage}.
The ejection event appeared to be one-sided, since only on MJD\,53817
do we see any evidence for a receding component.

\begin{figure}
\begin{center}
\includegraphics[width=0.45\textwidth,angle=0,clip=]{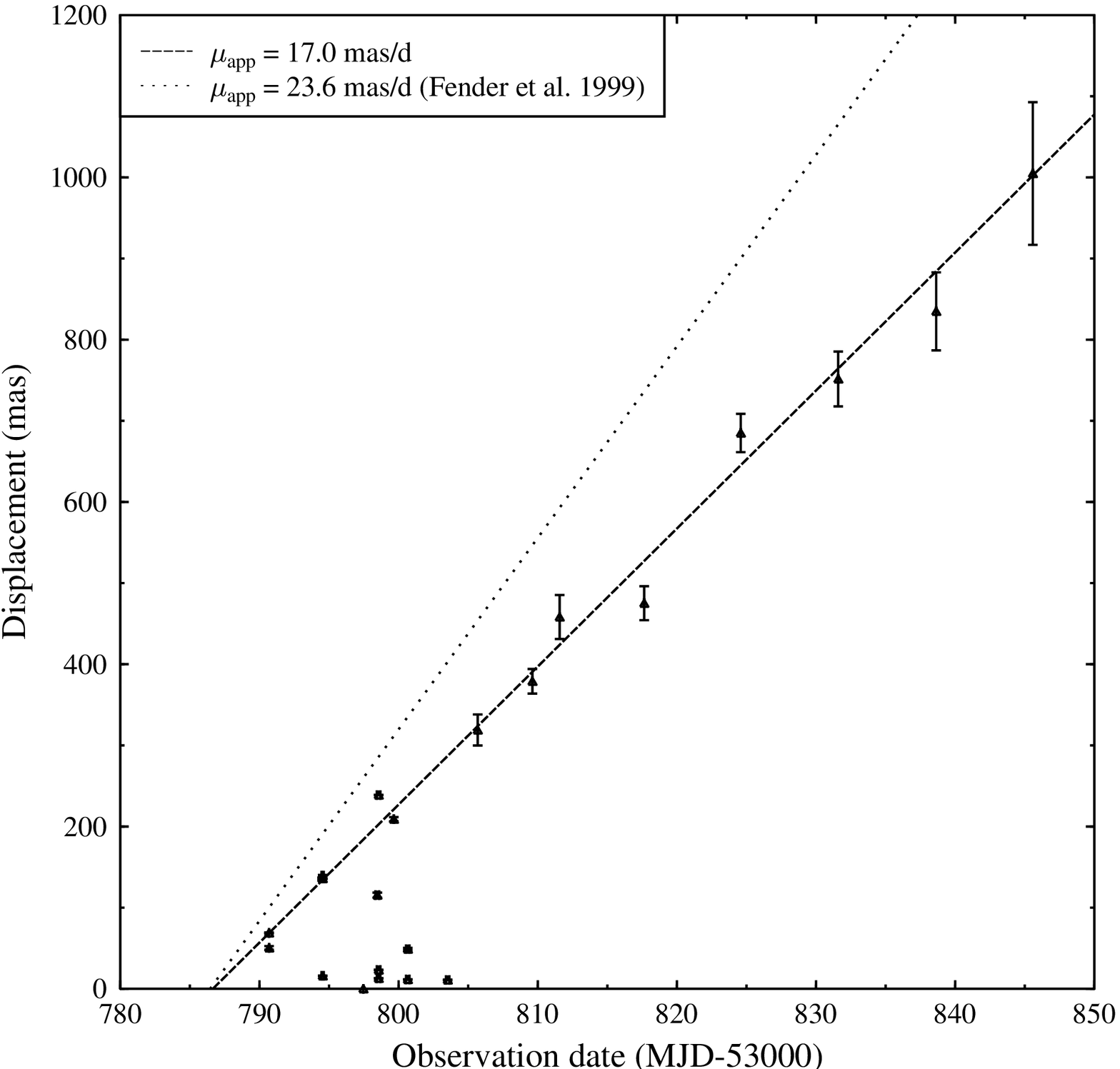}
\caption{Measured angular separations from the core of the SE
  component ejected on MJD~53786.  The separations measured with the
  VLBA and MERLIN are also included, which form the scatter of points at
  very low angular separations between MJD\,53794 and 53803.}
\label{fig:angsep}
\end{center}
\end{figure}

Tracking the ejected knot over such a long time interval gave
extremely good constraints on its proper motion, which we fitted as
$17.0\pm0.2$\,mas\,d$^{-1}$.  This fit (to the VLA proper motions
only) is shown in Fig.~\ref{fig:angsep}.  It is clearly incompatible
with the proper motions measured by \citet{Fen99}, and there is no
evidence for any deceleration.  To verify this, we fitted the angular
separations with both a straight-line fit (ballistic motion) and a
quadratic fit (corresponding to deceleration), and performed an F-test
\citep[e.g.][]{Pfe05}.  This gave a value $F_{1,10}=0.81$, implying
that the quadratic fit was not significantly better, i.e.\ there was
no significant evidence for any deceleration of the ejecta on the
angular scales probed by the VLA.

\subsubsection{Flux density decay}
\label{sec:flux_density}
\begin{figure}
\begin{center}
\includegraphics[width=0.45\textwidth,angle=0,clip=]{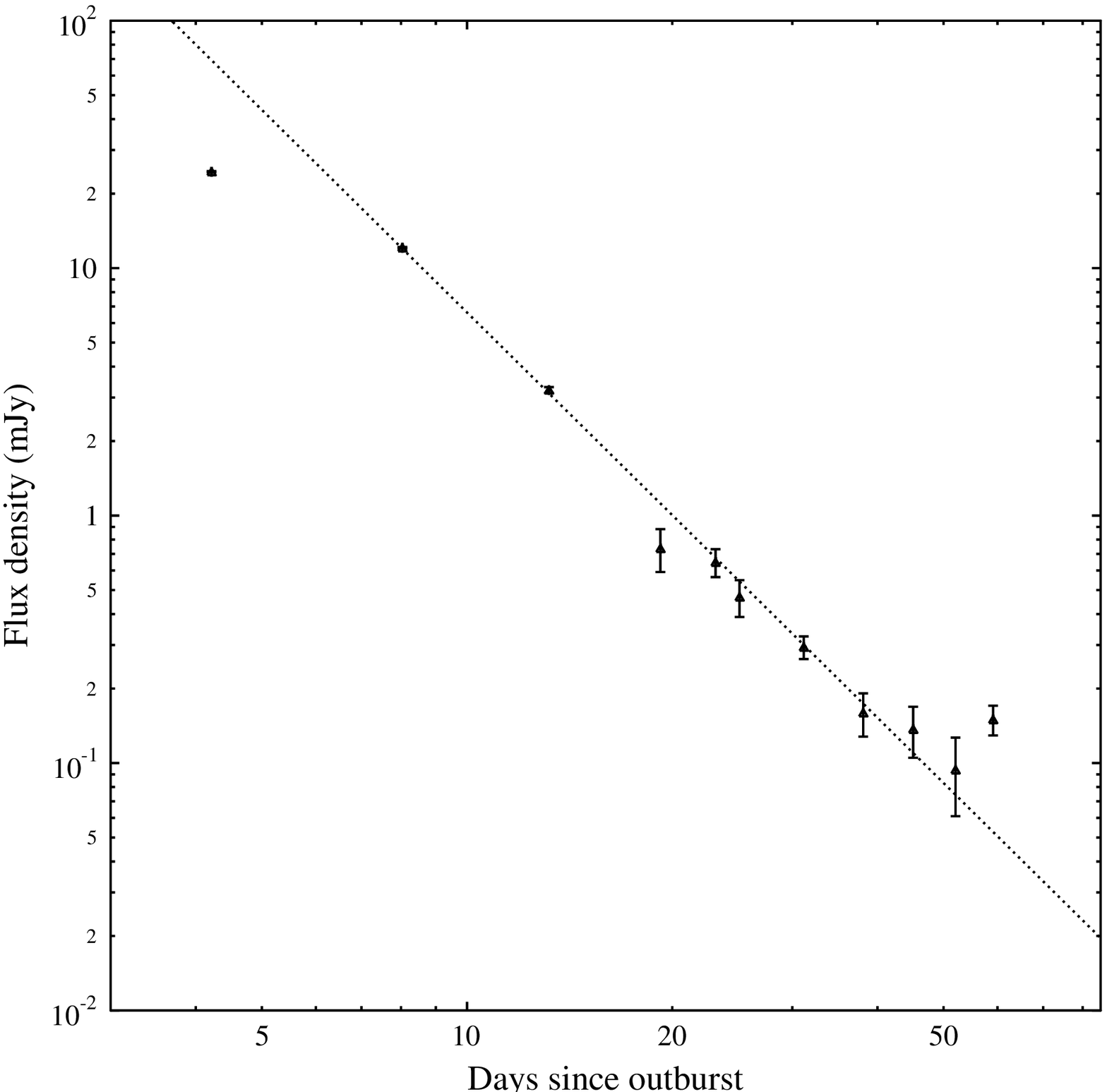}
\caption{Measured VLA flux densities of the approaching jet knot
  ejected on MJD\,53786, with a best-fitting line corresponding to a
  power-law decay of index 2.70 overlaid.  Measurements at frequencies
  other than 8.4\,GHz have had their flux densities scaled by an
  assumed spectral index $\alpha=-0.75$
  ($S_{\nu}\propto\nu^{\alpha}$).}
\label{fig:flux_densities}
\end{center}
\end{figure}

Between 8 and about 50 days after the 2006 February ejection event,
the flux density of the southeastern component appeared to follow a
power-law decay with time (and, since the motion is ballistic, then
also with angular separation), $S = S_0(t-t_0)^{-\tau}$, with the
exponent fitted as $\tau=2.70\pm0.05$, and where $t$ and $t_0$ are
both measured in days.  This is shown in
Fig.~\ref{fig:flux_densities}.  Where the observations were taken at a
lower frequency than 8.4\,GHz, the flux densities were extrapolated to
give the expected 8.4\,GHz flux density, assuming a steep synchrotron
spectrum, $S_{\nu} = S_0(\nu/\nu_0)^{-0.75}$.  The derived value of
$\tau$ is consistent with that of $2.6\pm0.5$ found by \citet{Rod99}
in GRS\,1915+105 on angular scales greater than 1\,arcsec from the
core.  In a previous outburst, they found $\tau=1.3\pm0.2$ on angular
scales within 1\,arcsec, and suggested that there could be a
transition from a regime of slowed to free expansion at some distance
from the core, as suggested for the case of SS\,433 by \citet{Hje88}.
Our observations track the knot out to a distance of $\sim1$\,arcsec,
suggesting that any transition from slowed to free expansion took
place much closer to the core during our outburst than in those
monitored by \citet{Rod99}.  Indeed, if we interpret the low flux
density of the first measurement in Fig.~\ref{fig:flux_densities} as
due to an initial phase of slowed expansion, then the transition to
free expansion would occur within 135\,mas.

\citet{Hje88} showed that the flux density decay for a single
impulsive ejection, expanding according to a conical jet model, would
scale as 
\begin{equation}
S \propto (t/t_0)^{-(7p-1)/(6+6\xi)}, \label{eq:fluxdecay}
\end{equation}
where $p$ is the index of the electron energy spectrum, and $\xi=1$
for slowed expansion and $\xi=0$ for free expansion.  Their derived
value of $p=2.32$ for SS\,433 gave indices of 1.27 during the slowed
expansion phase and 2.54 during free expansion, remarkably consistent
with the values found for GRS\,1915+105 by \citet{Rod99} and
suggesting a very similar electron energy spectrum in the two sources.
Our derived power-law index $\tau$ would in this case suggest that we
were observing the knot during its freely-expanding phase.  In fact,
from our fitted value $\tau=2.70\pm0.05$, we can calculate from Equation
\ref{eq:fluxdecay} a corresponding electron index $p=2.46\pm0.04$ and
a spectral index $\alpha=-0.73\pm0.02$.  While not strictly within the
error bars, this is broadly consistent with the previously-measured
values of 0.8 \citep{Fen99} and $0.84\pm0.03$ \citep{Mir94}, implying
that the data can be fairly well described by a freely-expanding
conical jet model with a single impulsive ejection event.

Alternatively, if the knot were to decelerate owing to interactions
with the ISM, the release of energy would be expected to cause a
brightening of the knot, as seen in both XTE J\,1550-564 \citep{Cor02}
and XTE J\,1748-288 \citep{Hje98}. The low initial point could thus be
due to a slight rebrightening between 4 and 8 days after the outburst
(i.e.\ within 135\,mas of the core), but with only a single point,
this remains highly speculative.  There is no evidence for any
significant rebrightening between 8 and 45 days after ejection, which
would argue against deceleration during that time period.  This is
consistent with a picture in which the jet was initially confined by a
dense medium close to the source, with which it interacted.  Once the
angular separation was sufficiently great, the lower density would
cause the transition from slowed to free expansion, and in the absence
of sufficient material to cause deceleration, the knot then moved
outwards ballistically.  However, with such sparse sampling of the
early phase of the outburst, this scenario cannot be confirmed with
the available data.

\subsection{Combining the data}
There is only one epoch of VLBA data (MJD\,53794) in which the
approaching component seen with the VLA was resolved.  Subsequently,
it is to be assumed that the knot had moved too far from the core, and
expanded and faded too much to be detectable with the VLBA.  For that
one set of overlapping data, the two arrays detected the component at
almost identical positions, and the VLBA saw only a single component,
as opposed to a string of successive ejecta.  This would suggest that
it is unlikely that resolution effects between the different arrays
are causing the measurement of different proper motions.  MERLIN did
not at any stage detect the original ejecta, but the first epoch of
MERLIN data is later still, MJD\,53797.

It is also not possible to track any of the VLBA knots between VLBA
epochs, so we cannot definitively assign a proper motion to any of
these knots.  We strongly encourage future high-resolution observations
to begin observing the source as quickly as possible after a flare, in
order to be able to track the knots before they fade.

Nevertheless, we can use the start of the flare detected in the first
epoch of MERLIN data to constrain the ejection event giving rise to
the knots seen in the second epoch of VLBA data.  If the outer pair of
knots in the VLBA image are assumed to have been ejected at the time
the flux density started to rise, it would imply proper motions of
$21.1\pm0.3$\,mas\,d$^{-1}$ for the southeastern (approaching) knot and
$11.7\pm0.3$\,mas\,d$^{-1}$ for the northwestern (receding) knot,
closer to what has previously been observed by the VLBA and
MERLIN.  But the start of the MERLIN flare is the latest possible time
when the ejection event could have occurred.  It is possible that the
ejection happened earlier, moved outwards, and only caused an increase
in the total flux density once it reached a point at which the knot
became optically thin at the observing frequency.  Thus these
observations do not conclusively rule out an earlier ejection date and
consequently a lower proper motion.

\subsubsection{The receding component}
\label{sec:receding}
The receding component was in no case unambiguously detected in our
data from the 2006 February outburst, although there are hints of a
receding component in the image of MJD\,53817.  We note that during
previous ejection events, the receding component has not always been
detected.  \citet{Rod99} observed the receding component in four of
their five reported events, \citet{Fen99} detected it in one of their
three events, and \citet{Mil05} in two of their four events.  We
assume that the outburst was not sufficiently bright to detect the
Doppler-deboosted receding component once it was significantly far out
to be resolved from the core.

The first epoch of VLBA data shows evidence for ejecta at a position
angle consistent with being the receding northwestern component (see
Table~\ref{tab:vlba_measurements}).  The 2.3\,GHz flux density peak is
significantly closer to the core than that at 8.4\,GHz, possibly
suggesting that there were higher-energy particles present at the
leading edge of the ejecta, consistent with the internal shock
interpretation.  Receding components are also seen at much smaller
angular separations in the second epoch.  We can use the angular
separations from the core of corresponding approaching and receding
knots to constrain the value of $\beta\cos\theta$ for those particular
ejections, via
\begin{equation}
\beta\cos\theta = \frac{\mu_{\rm app}-\mu_{\rm rec}}{\mu_{\rm
    app}+\mu_{\rm rec}},
\label{eq:bct}
\end{equation}
where the time dependence of $\mu$ cancels out since we see all
components in a single image.  Applying this approach to the images of
the first epoch (MJD\,53794.5), this gives $\beta\cos\theta =
0.33\pm0.01$ at 8.4\,GHz and $0.46\pm0.02$ at 2.3\,GHz, close to the
previous estimates of $0.323\pm0.011$ \citep{Mir94} and $0.41\pm0.02$
\citep{Fen99}.  Assuming that the two bright northwestern (receding)
components seen in the second epoch of VLBA data (MJD\,53798.6)
correspond to the two bright approaching components, we find
$\beta\cos\theta = 0.27\pm0.01$ for the furthest out pair (SE2 and NW2
in Table \ref{tab:vlba_measurements}) and $0.22\pm0.03$ for the other
pair (SE1 and NW1).  In both cases this is significantly smaller than
the values found in the literature, suggesting a possible
misdentification of corresponding components or of the true core
location.

\begin{figure}
\begin{center}
\includegraphics[width=\columnwidth,angle=0,clip=]{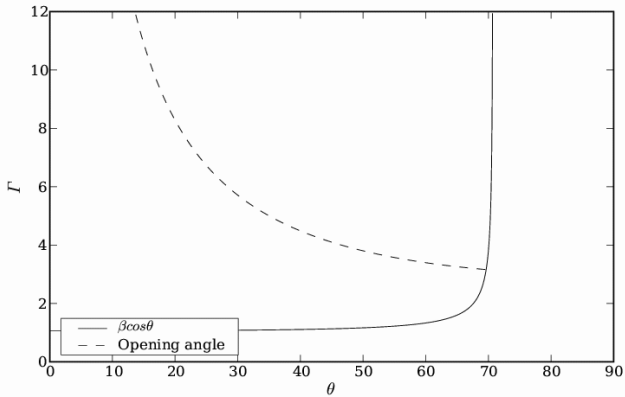}
\caption{Variation of the derived Lorentz factor for the approaching
    jet knot with inclination angle to the line of sight.  {\it Dotted
    line:} Minimum possible Lorentz factor derived from the measured
    upper limit on the jet opening angle, assuming free expansion.  {\it
    Solid line:} Bulk Lorentz factor derived from the measured value
    of $\beta\cos\theta$ from the first epoch of VLBA observations.}
\label{fig:opening_angles}
\end{center}
\end{figure}

If, as suggested in Section \ref{sec:flux_density}, the jet is freely
expanding, then observational constraints on the opening angle and the
derived value of $\beta\cos\theta$ can be used to place a lower limit
on the Lorentz factor of the jet knot, as shown by Miller-Jones,
Fender \& Nakar (2006).  The last 8.4-GHz VLA observation (MJD
53838.6) gives us the best constraints on the opening angle of the
(unresolved) jet knot, $\phi<19.6^{\circ}$.  Together with the
constraint on $\beta\cos\theta$ from the first set of 8.4-GHz VLBA
images (Section \ref{sec:receding}), this allows us to derive a lower
limit to the bulk Lorentz factor for any given value of the
inclination angle, via
\begin{equation}
\Gamma = \left(1+\frac{\beta_{\rm exp}^2}{\tan^2\phi\sin^2
\theta}\right)^{1/2},
\end{equation}
where $\beta_{\rm exp}c$ is the jet knot expansion speed.  The derived
Lorentz factors for free expansion ($\beta_{\rm exp}=1$) are shown in
Fig.~\ref{fig:opening_angles}.  This suggests that the Lorentz factor
should be $>3.2$ during the freely-expanding phase.  In order to agree
with the constraint from the observed value of $\beta\cos\theta$, the
source must be close to $d_{\rm max}$ (the maximum distance allowed by
the measured proper motions of the radio jets, defined as
$c/\sqrt{\mu_{\rm app}\mu_{\rm rec}}$).  Such a Lorentz factor would
agree with those derived by \citet{Fen99} from MERLIN observations,
assuming a distance close to $d_{\rm max}=11.2$\,kpc.

\subsubsection{Convolutions}
\label{sec:convolutions}

For the few epochs when we have overlapping observations with
different arrays, it is possible to convolve the higher resolution
data to the resolution of the more compact array in order to check for
consistency.  Convolving the VLBA data to the VLA resolution on
MJD\,53794.5, the resulting source morphology was found to be very
similar.  While not altogether unexpected, since the VLBA had detected
a single extended component at the same position as the VLA detection,
this is reassuring.  Convolving the VLBA data to the MERLIN resolution
on MJD\,53798.6, there was no evidence in the VLBA data at either
frequency for the approaching component seen with MERLIN.  This would
tend to suggest that the component seen in the MERLIN data was
relatively diffuse, since the largest angular scale probed by the VLBA
is 41\,mas at 8.4\,GHz and 147\,mas at 2.3\,GHz, whereas MERLIN probed
angular scales in the range 47\,mas to 1.88\,arcsec.  This effect
could also have prevented the high-resolution arrays from seeing the
original ejected component after MJD\,53794; once it had expanded
sufficiently such that there was no structure on scales small enough
to be seen with the high resolution arrays, it could not be detected.
Coupled with the greater sensitivity of the VLA, this explains the
non-detection of the original southeastern component in the last three
VLBA epochs and the last MERLIN epoch.

An attempt was also made to convolve the VLBA data taken on
MJD\,53798.6 and 53800.7 to the VLA resolution on MJD\,53799.7.  Since
the observed VLBA structure was fairly small-scale and compact, there
was no detection of the extension seen in the VLA image.  The proper
motion discrepancy cannot therefore be definitively ascribed to the
spatial averaging of underlying compact structures with the VLA.
\citet{Hje95} pointed out that a combination of the flux density of
the ejecta fading with time and beam smearing effects could lead to
inaccurate proper motion estimates.  While this cannot be ruled out
with the present data, the images of MJD\,53794 suggest that this
effect is not occurring in the current datasets, since only a single
set of VLBA ejecta was observed, rather than a continuous string of
knots.

\section{Discussion}
\label{sec:discussion}
The fact that we have measured the same proper motion (to within
errors) as seen with the VLA in 1994 suggests that there has been no
lasting fundamental change in the system since that time.  This would
seem to rule out the secular change scenario postulated by
\citet{Mil05} to explain the proper motion discrepancy.  If, as
suggested, the radio jets had only switched on in 1992, since when
they had evacuated a cavity in the interstellar medium (ISM) allowing
the jets to propagate further at high velocity before decelerating,
then a high proper motion, of order 23.6\,mas\perday, should also be
measured with our VLA data, which was not the case.

The VLBA data would suggest that resolution effects are not
responsible for the discrepancy between the proper motions measured on
large and small angular scales.  As well as the range of angular
scales probed by MERLIN and the VLA overlapping, the image of
MJD\,53794 shows only a single VLBA component, rather than a string of
successive knots.  The convolutions presented in Section
\ref{sec:convolutions} further argue against this possibility.

From the range of proper motions that we have measured, and from the
values quoted in the literature (Table \ref{tab:propermotions}), it
would appear that GRS\,1915+105 is capable of producing ejecta with a
range of proper motions.  In no case has there ever been any evidence
for deceleration.  Thus, if the spread of proper motions is intrinsic,
this would imply that the speeds of the jet knots are not fixed by the
properties of the black hole powering the jets (its mass, spin or
magnetic field).  The variation in the speed of the ejecta is in fact
a feature of the so-called ``unified model'' of black hole X-ray
binaries proposed by Fender, Belloni \& Gallo (2004), whereby the
Lorentz factor of the jets increases as the X-ray spectrum softens in
the Very High State, leading to internal shocks within the flow which
are seen as discrete ejecta.  For those discrete ejecta to have
different observed proper motions would then imply a variable velocity
discrepancy between the jet Lorentz factor in the hard and quiescent
states, and that in the soft state.  Since the inner edge of the
accretion disc is believed to move inwards during this transition
between the hard and soft states, this would imply a different inner
disc radius for each outburst.  This is certainly plausible, and would
also explain the variable values of $\beta\cos\theta$ quoted in
Section \ref{sec:receding} and in the literature.

\begin{figure}
\begin{center}
\includegraphics[width=\columnwidth,angle=0,clip=]{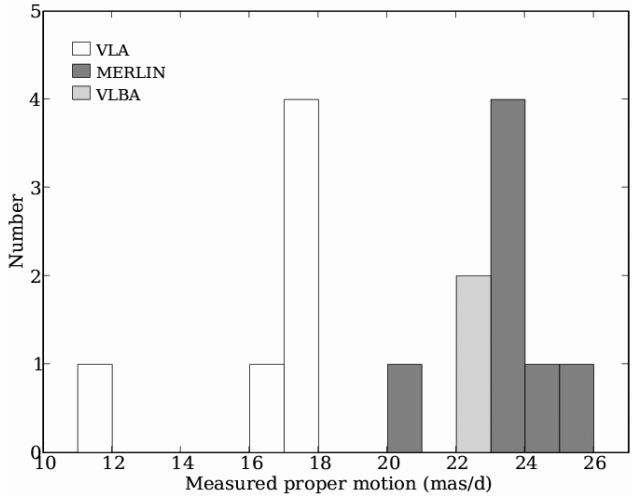}
\caption{Histogram of the measured proper motions of the approaching
  jet knot of GRS\,1915+105, colour-coded by array.  Data taken from
  Table \ref{tab:propermotions}.  The measurement of
  $11\pm2$\,mas\perday\ is considered unreliable, since the jets were
  resolved in only one image.}
\label{fig:mu_distribution}
\end{center}
\end{figure}

There is in fact no strong evidence for the jets in any X-ray binary
to have identical proper motions during different outbursts.  The
number of systems in which radio proper motions from multiple
outbursts have been measured is small (GRS\,1915+105, Cygnus X-3, GRO
J\,1655-40, and SS\,433).  SS\,433 has the most well-monitored and
constant jet velocity of $\sim0.26c$, but \citet{Blu05} have shown
that the jet velocity does indeed vary, with a standard deviation of
0.013 in $\beta$.  Thus while a varying proper motion in GRS\,1915+105
would not necessarily go against the trend, it seems highly
coincidental that if the observed speeds are drawn from some
underlying distribution, there should be such a clear distinction
between the high proper motions observed only by the high-resolution
arrays, and the lower values seen only with the VLA (see
Fig.~\ref{fig:mu_distribution}).  At present however, we are still in
the regime of small number statistics.  Nevertheless, applying a
Kolmogorov-Smirnov test \citep[e.g.][]{Pre92} to see whether the VLA
and higher-resolution observations could be drawn from the same
distribution gives a probability of $3.63\times 10^{-4}$ ($8.24\times
10^{-4}$ if the August 1995 point is ignored).  This therefore seems
unlikely.

\begin{figure*}
\begin{center}
\includegraphics[width=\textwidth,angle=0,clip=]{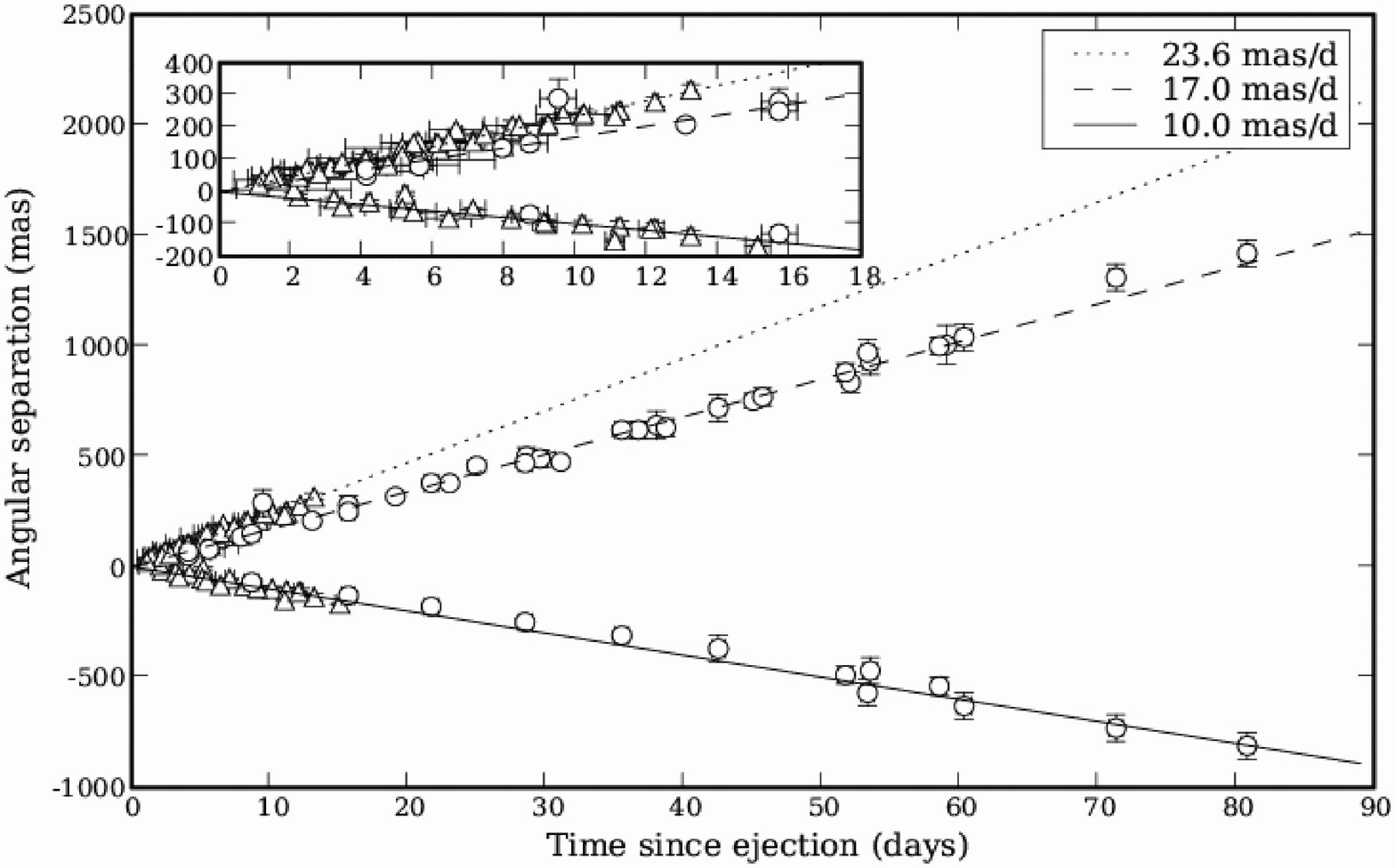}
\caption{Variation in measured angular separation with time since
  ejection. Data taken from VLA \citep[][and Table
  \ref{tab:vla_measurements}]{Rod99}, MERLIN \citep{Fen99,Mil05} and
  VLBA \citep[][and Table \ref{tab:vlba_measurements}]{Dha00}
  observations of GRS\,1915+105.  Ejection dates were taken from the
  ballistic proper motion fits reported in the source literature.
  Open circles show VLA data, open triangles show VLBA or MERLIN data.
  The inset shows a zoomed-in version of the smallest angular scales.
  There is no clear evidence for deceleration on any angular scale.}
\label{fig:all_obs}
\end{center}
\end{figure*}

Since we were unable to track the ejecta from the initial event with
the VLBA and MERLIN, we cannot definitively rule out the possibility
that the ejecta decelerated on angular scales $<100$\,mas, before the
VLA was able to accurately resolve them.  In fact, extrapolating the
proper motion measured with the VLA back to zero angular separation
gives an ejection date of MJD\,53786.65, 1.86\,d prior to the peak of
the 15-GHz radio flare detected with the Ryle Telescope.  Although
since, as previously noted, the radio flare is the latest possible
date of ejection such that the true ejection date could have been
somewhat earlier, this might be taken as suggestive of deceleration
between the radio flare and the first set of VLA observations.
Assuming an initial proper motion of 23.6\,mas\perday\ and an ejection
date corresponding to the time of the 15-GHz radio flare, then
extrapolating the best fitting line for the VLA proper motion
backwards in time would imply that the deceleration would be occurring
on angular scales of $\sim70$\,mas from the core.  Previous MERLIN
observations \citep{Fen99,Mil05} have tracked the ejecta out to
angular scales of $>300$\,mas, with no compelling evidence for
deceleration.  Fig.~\ref{fig:all_obs} shows how the angular
separations changed with time since ejection for all the events
reported in the literature (using the derived ejection dates).
From this figure, there is no clear evidence that deceleration occurs
at any specific distance from the core.  The diagram shows two clear
tracks, one sampled by the higher-resolution arrays, and the other,
with lower proper motions, by the VLA.  The receding jet angular
separations all appear to lie on the same track.

Circumstantial evidence for deceleration on small scales in
GRS\,1915+105 can be seen in the linear polarisation `rotator event'
observed in January 2001 by \citet{Fen02}.  The electric vector
position angles measured with the Australia Telescope Compact Array
(ATCA) at 4.80 and 8.64\,GHz rotated smoothly together through
50\degr\ over the course of $\sim0.2$\,d.  The constant separation
between the position angles at the two frequencies ruled out Faraday
rotation, and this was attributed to a changing magnetic field
orientation.  We note however that \citet{Bla79} suggested that an
accelerating (or decelerating) jet would show a swing in polarisation
position angle as the aberration angle varied.  If this was the true
explanation for the `rotator event', and we attribute the detected
polarisation to the southeastern (approaching) component \citep[as
seen by][]{Fen99,Mil05}, which we assume to have been ejected at the
time of the spike in the {\it RXTE} data shown by \citet{Fen02}, then
this deceleration would have been occurring somewhere between 0.2 and
1.0\,d after ejection.  Such an effect was not seen in the ATCA
observations of the March 2001 ejection event, taken between 1.5 and
1.9\,d after the zero-separation date derived by \citet{Mil05}.  This
would imply that if all outbursts followed the same pattern,
deceleration would have to occur within $\sim50$\,mas.  Polarisation
position angles measured by MERLIN \citep{Fen99,Mil05} show large
swings (albeit at a single frequency, such that Faraday rotation
cannot be discounted) between 3 and 4\,d after ejection, which would
constrain the deceleration to occur within 70--100\,mas.  However,
this evidence remains circumstantial, and would appear to be at odds
with the MERLIN points 8--13\,d after ejection on the
23.6\,mas\,d$^{-1}$ track in the inset to Fig.~\ref{fig:all_obs}.

If the jets are indeed decelerating on angular scales of order
70\,mas, then for a given source distance, we can find the change in
$\beta$, and hence the change in the bulk Lorentz factor $\Gamma$ and
the amount of energy lost to the environment.

The proper motion of an approaching jet knot is given by
\begin{equation}
\mu_{\rm app} = \frac{c\beta\sin\theta}{d(1-\beta\cos\theta)},
\label{eq:mu_app}
\end{equation}
where $\beta$ is the intrinsic jet speed, $\theta$ is the inclination
angle of the jet axis to the line of sight, $c$ is the speed of light,
and $d$ is the source distance.  For a constant source distance, a
changing proper motion requires either a change in $\beta$ or a change
in $\theta$.  The value of the product $\beta\cos\theta$ is
constrained by measurements of the approaching and receding proper
motions, via Equation~\ref{eq:bct}.  This was measured by
\citet{Fen99} to be $\beta\cos\theta = 0.41\pm0.02$ for the
MERLIN-scale jets (with $\mu_{\rm app} = 23.6\pm0.5$\,mas\perday) and
by \citet{Mir94} to be $\beta\cos\theta = 0.323\pm0.011$ for the
VLA-scale jets ($\mu_{\rm app}=17.6\pm0.4$\,mas\perday).  Combining
Equations \ref{eq:bct} and \ref{eq:mu_app} gives a value for $\beta$,
\begin{equation}
\beta = \sqrt{\frac{\mu_{\rm app}^2 d^2
    (1-\beta\cos\theta)^2}{c^2}+(\beta\cos\theta)^2},
\end{equation}
taking $\beta\cos\theta$ to be a measured quantity.  For a given
distance, we can solve for the jet speed, $\beta$, and hence find the
bulk Lorentz factor $\Gamma=(1-\beta^2)^{-1/2}$.

\begin{figure}
\begin{center}
\includegraphics[width=\columnwidth,angle=0,clip=]{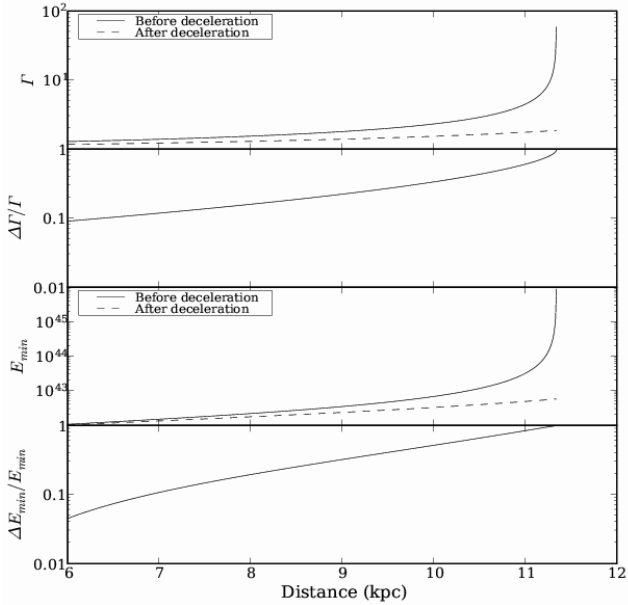}
\caption{{\it Top panel:} variation with distance of Lorentz factor prior to
  deceleration ($\mu_{\rm app}=23.6\pm0.5$\,mas\perday,
  $\beta\cos\theta = 0.41$) and after deceleration ($\mu_{\rm
  app}=17.6$\,mas\perday, $\beta\cos\theta = 0.323$).  {\it Second
  panel:} Fractional change in Lorentz factor during deceleration.
  {\it Third panel:} Minimum energy calculated before and after
  deceleration, using the Lorentz factors shown above.  {\it Bottom
  panel:} Fractional change in minimum energy of the jet knots during deceleration.}
\label{fig:calcs}
\end{center}
\end{figure}

Further, we can find the minimum energy associated with the
synchrotron emission from the jet knots, given by
\citep[e.g.][]{Lon94}
\begin{equation}
E_{\rm min} =
\frac{7}{6\mu_0}V^{3/7}\left(\frac{3\mu_0}{2}G(\alpha)\eta
  L_{\nu}\right)^{4/7},
\end{equation}
where $V$ is the source volume, $\mu_0$ is the permeability of free
space, $(\eta-1)$ is the ratio of the energy in protons to that in
relativistic electrons, $L_{\nu}$ the source luminosity, and
\begin{equation}
G(\alpha) \propto \left(\nu_{\rm min}^{\alpha+1/2}-\nu_{\rm
  max}^{\alpha+1/2}\right)\nu_{\rm obs}^{-\alpha}.
\label{eq:g_alpha}
\end{equation}
Assuming a spherical source of radius given by the light crossing time
$\Delta t$, we can express $V = 4\pi (c\Delta t)^3/3$.  The luminosity
is $L_{\nu} = 4\pi d^2 S_{\nu}$, where $S_{\nu}$ is the
source flux density.  Factoring in the relevant constants of
proportionality (assuming $\alpha = -0.75$; a different value would
only change the numerical constant), we find that
\begin{multline}
E_{\rm min}(\Gamma = 1) = 3.5\times10^{33}\eta^{4/7} \left(\frac{\Delta t}{{\rm
    s}}\right)^{9/7} \left(\frac{d}{{\rm kpc}}\right)^{8/7}\\
    \times\left(\frac{\nu}{{\rm GHz}}\right)^{2/7} \left(\frac{S_{\nu}}{{\rm
    mJy}}\right)^{4/7} \qquad {\rm erg},
\label{eq:e_min}
\end{multline}
where the rise time of the flare $\Delta t$, the flux density
$S_{\nu}$, and the frequency $\nu$ at which the flare is observed
should be measured in the source rest frame, equivalent to the
observer's frame for $\Gamma=1$.  If the source is moving
relativistically, then the transforms $\nu = \delta\nu^{\prime}$,
$S_{\nu} = \delta^{3-\alpha}S_{\nu}^{\prime}$, and $\Delta t =
\delta^{-1}\Delta t^{\prime}$ must be applied, where $\delta$ is the
Doppler factor $[\Gamma(1-\beta\cos\theta)]^{-1}$.  Care must be taken
when transforming Equation \ref{eq:g_alpha}, since the standard
procedure is to set $\nu_{\rm max}=\infty$ and $\nu_{\rm obs} = \nu_{\rm
min}$.  This should however be done after boosting, since the
transformation of $\nu_{\rm obs}$ is already accounted for by the
extra factor $\alpha$ in the transformation of $S_{\nu}$.  Thus only
$\nu_{\rm min}$ should be relativistically boosted.  Taking this into
account, and multiplying the result by an extra factor $\Gamma$ to
account for the energy associated with the bulk motion of the flow, the
appropriate correction is then
\begin{equation}
E_{\rm min}(\Gamma\neq 1) = \Gamma\delta^{-5/7}E_{\rm min}(\Gamma=1),
\end{equation}
if the measured values in the observer's frame are to be used.  Taking
the values quoted by \citet{Fen99} of $\alpha = -0.8$, $\Delta t$ =
12\,h, $\nu = 2.3$\,GHz and $S_{\nu}=550$\,mJy, together with
$\eta=1$, then for a given distance, calculating $\Gamma$ (using
$\beta\cos\theta=0.41$ and $\mu_{\rm app}=23.6$\,mas\perday) allows us
to find the energy of the jet knots.  Subsequently allowing a
deceleration to $\beta\cos\theta=0.323$ and $\mu_{\rm
app}=17.6$\,mas\perday\ allows us to calculate the fractional energy
loss from such a deceleration.  The results of these calculations are
shown in Fig.~\ref{fig:calcs}.
If the source is decelerating, then if it lies close to $d_{\rm max}$,
it would lose almost all its energy in decelerating from 23.6 to
17.6\,mas\,\perday.  Even at a distance of 6\,kpc, its bulk Lorentz
factor would decrease by 10 per cent and it would deposit 4 per cent
of its energy into the surrounding ISM.

The only X-ray binaries in which conclusive evidence for deceleration
has thus far been observed are XTE J\,1748-288 and XTE J\,1550-564.
Four years after the initial ejection event of XTE J\,1550-564 in
September 1998 \citep{Han01}, the jets were seen to reappear at radio
and X-ray wavelengths \citep{Cor02} at angular separations of
$\sim23^{\prime\prime}$.  Proper motions were obtained for the X-ray
knots \citep{Kaa03} which showed that the knots were decelerating,
presumably due to some interaction with the surrounding ISM.  Similar
X-ray knots have also been detected in H\,1743-322 \citep{Cor05}, and
while deceleration has not been directly observed, it is possible that
these knots were also powered by bulk deceleration.  It was postulated
that the jet knot from the 1998 outburst of XTE J\,1748-288
decelerated \citep{Hje98}, although no definitive analysis is
available in the literature.  The knot was seen to stop at an angular
separation of $\sim300$\,mas and brighten, almost as if it had hit a
wall.  Another case where there is possible evidence for deceleration
is that of Cygnus X-3, where the milliarcsecond-scale structure shows
a brighter, approaching southern jet \citep{Mio01,Mil04} and a fainter
receding northern counterjet, whereas the arcsecond-scale structure
shows the northern jet knot to be brighter and at a larger angular
separation from the core \citep{Mar01}.  This could also be
interpreted as evidence for deceleration between these two sets of
angular scales.  Thus while deceleration has been directly observed on
large angular scales, XTE J\,1748-288 and Cygnus X-3 are the only
cases where this might be occurring so close to the centre of the
system as required here in the case of GRS\,1915+105.  Cygnus X-3,
while a very active radio emitter, has a Wolf-Rayet companion
\citep{van96} with a dense stellar wind, so its immediate environment
would likely be denser than that surrounding GRS\,1915+105, which has
a K-M \textsc{III} companion \citep{Gre01}.  There is also still
speculation about the nature of the compact object in Cygnus X-3.
Little is known about XTE J\,1748-288.  Its compact object is very
likely to be a black hole \citep{Rev00}, but the nature of its
companion is as yet unknown.  However, in this system deceleration was
clearly observed, with the jet knot appearing to stop and brighten,
which is certainly not the case in GRS\,1915+105.

Ultimately, the evidence for deceleration in GRS\,1915+105 is not
conclusive.  It is perhaps most likely that the jet speed in this
system is indeed variable, although the VLA and higher-resolution
proper motion measurements do not appear to be drawn from the same
distribution of speeds.  More observations are certainly needed to
verify this, since we are still in the regime of small number
statistics.

\section{Conclusions}
\label{sec:conclusions}
We have measured the proper motion of a jet knot ejected during the
2006 February outburst of GRS\,1915+105 as 17.0\,mas\perday.  This is
the first measurement of such a low proper motion since 1995, and
shows that there has been no significant permanent change in the
system which would allow the jet knots to propagate at a faster speed.
The VLBA images have shown that although the first ejection was the
brightest, there were in fact several ejection events over the course
of two weeks after the radio outburst was first detected.  The
brightest jet knot shows no evidence for deceleration on angular
scales greater than 100\,mas, and its flux density decays as a simple
power law between 8 and 45 days after the ejection event, with a power
law index suggestive of a freely-expanding jet.  In this case, the
bulk Lorentz factor of the jet would be constrained to $\Gamma>3$.  By
making simultaneous observations with multiple arrays probing
different angular scales, we showed that the slower proper motions
measured by the VLA are unlikely to be due to resolution effects.

Despite the aim of this set of observations being to cover a single
outburst over a large range of angular scales using the VLBA, MERLIN
and the VLA, the high-resolution observations could not be triggered
early enough to properly track the knots seen with the VLA, and we
were unable to conclusively resolve the discrepancy in the proper
motions.  In order to find out what is really going on, it will be
important to begin observing within one or two days of the detection
of the outburst with the VLBA and MERLIN, to track the knots on the
smallest angular scales while they are still bright, and to attempt to
follow the same knots with the A-configuration VLA as they move
outwards.  Following a brighter outburst would also help, enabling
monitoring with higher positional accuracy and out to larger angular
distances (particularly important for the high-resolution arrays).
If, as suggested by \citet{Tru06}, the accretion disc in GRS\,1915+105
is now almost empty, then the source may soon switch off.  There may
therefore be few opportunities left to initiate such a monitoring
programme, since it requires a bright outburst to occur while the VLA
is in its A configuration.

\section*{Acknowledgments}
The National Radio Astronomy Observatory is a facility of the National
Science Foundation operated under cooperative agreement by Associated
Universities, Inc.  The X-ray data presented are quick-look results
provided by the ASM/{\it RXTE} team.  MERLIN is operated as a National
Facility by the University of Manchester at Jodrell Bank Observatory
on behalf of the Particle Physics and Astronomy Research Council
(PPARC).  JCAM-J would like to thank Sebastian Jester for useful
discussions and the anonymous referee for their constructive feedback.
\label{lastpage}

\bibliographystyle{mn2e}

\begin{thebibliography}{}

\bibitem[Blandford \& Konigl(1979)]{Bla79}
Blandford R.~D., Konigl A., 1979, ApJ, 232, 34

\bibitem[Blundell \& Bowler(2005)]{Blu05}
Blundell K.~M., Bowler M.~G., 2005, ApJ, 622, L129

\bibitem[Castro-Tirado et al.(1992)]{Cas92}
Castro-Tirado A.~J., Brandt S., Lund N., 1992, IAU Circular, 5590, 2

\bibitem[Corbel et al.(2002)]{Cor02}
Corbel S., Fender R.~P., Tzioumis A.~K., Tomsick J.~A., Orosz J.~A.,
Miller J.~M., Wijnands R., Kaaret P., 2002, Science, 298, 196

\bibitem[Corbel et al.(2005)]{Cor05}
Corbel S., Kaaret P., Fender R.~P., Tzioumis A.~K., Tomsick J.~A.,
Orosz J.~A., 2005, ApJ, 632, 504

\bibitem[Dhawan et al.(2000)]{Dha00}
Dhawan V., Mirabel I.~F., Rodr{\'{\i}}guez L.~F., 2000, ApJ, 543, 373

\bibitem[Fender et al.(1999)]{Fen99}
Fender R.~P., Garrington S.~T., McKay D.~J., Muxlow T.~W.~B., Pooley
G.~G., Spencer R.~E., Stirling A.~M., Waltman E.~B., 1999, MNRAS, 304,
865

\bibitem[Fender et al.(2002)]{Fen02}
Fender R.~P., Rayner D., McCormick D.~G., Muxlow T.~W.~B., Pooley G.~G., Sault R.~J., Spencer R.~E., 2002, MNRAS, 336, 39

\bibitem[Fender et al.(2004)]{Fen04}
Fender R.~P., Belloni T.~M., Gallo E., 2004, MNRAS, 355, 1105

\bibitem[Greiner et al.(2001)]{Gre01}
Greiner J., Cuby J.~G., McCaughrean M.~J., Castro-Tirado A.~J.,
Mennickent R.~E., 2001, A\&A, 373, L37

\bibitem[Hannikainen et al.(2001)]{Han01}
Hannikainen D., Campbell-Wilson D., Hunstead R., McIntyre V., Lovell
J., Reynolds J., Tzioumis T., Wu K., 2001, Astrophysics and Space
Science Supplement, 276, 45 

\bibitem[Heinz(2002)]{Hei02}
Heinz S., 2002, A\&A, 388, L40

\bibitem[Hjellming \& Johnston(1988)]{Hje88}
Hjellming R.~M., Johnston K.~J., 1988, ApJ, 328, 600

\bibitem[Hjellming \& Rupen(1995)]{Hje95}
Hjellming R.~M., Rupen M.~P., 1995, Nature, 375, 464

\bibitem[Hjellming et al.(1998)]{Hje98}
Hjellming R.~M., Rupen M.~P., Mioduszewski A.~J., Smith D.~A., Harmon B.~A., Waltman E.~B., Ghigo F.~D., Pooley G.~G., 1998, Bulletin of the American Astronomical Society, 30, 1405

\bibitem[Kaaret et al.(2003)]{Kaa03}
Kaaret P., Corbel S., Tomsick J.~A., Fender R., Miller J.~M., Orosz J.~A., Tzioumis A.~K., Wijnands R., 2003, ApJ, 582, 945

\bibitem[Kaiser et al.(2004)]{Kai04}
Kaiser C.~R., Gunn K.~F., Brocksopp C., Sokoloski J.~L., 2004, ApJ, 612, 332

\bibitem[Longair(1994)]{Lon94}
Longair M.~S., 1994, High energy astrophysics. Vol.2: Stars, the galaxy and the interstellar medium. Cambridge: Cambridge University Press, 2nd ed.

\bibitem[Mart\'\i\ et al.(2001)]{Mar01}
Mart\'\i, J., Paredes, J.M., \& Peracaula, M.  2001, A\&A, 375, 476

\bibitem[Mioduszewski et al.(2001)]{Mio01}
Mioduszewski A.~J., Rupen M.~P., Hjellming R.~M., Pooley G.~G., Waltman E.~B., 2001, ApJ, 553, 766

\bibitem[Miller-Jones et al.(2004)]{Mil04}
Miller-Jones J.~C.~A., Blundell K.~M., Rupen M.~P., Mioduszewski
A.~J., Duffy P., Beasley A.~J., 2004, ApJ, 600, 368

\bibitem[Miller-Jones et al.(2005)]{Mil05}
Miller-Jones J.~C.~A., McCormick D.~G., Fender R.~P., Spencer R.~E.,
Muxlow T.~W.~B., Pooley G.~G., 2005, MNRAS, 363, 867

\bibitem[Miller-Jones et al.(2006)]{Mil06a}
Miller-Jones J.~C.~A., Fender R.~P., Nakar E., 2006, MNRAS, 367, 1432

\bibitem[Miller-Jones et al.(2006)]{Mil06}
Miller-Jones J.~C.~A., Rupen M.~P., Trushkin S.~A., Pooley G.~G.,
Fender R.~P., 2006, ATel 758

\bibitem[Mirabel \& Rodr{\'{\i}}guez(1994)]{Mir94}
Mirabel I.~F., Rodr{\'{\i}}guez L.~F., 1994, Nature, 371, 46

\bibitem[Pfenniger \& Revaz(2005)]{Pfe05}
Pfenniger D., Revaz Y., 2005, A\&A, 431, 511

\bibitem[Pooley \& Fender(1997)]{Poo97}
Pooley G.~G., Fender R.~P., 1997, MNRAS, 292, 925

\bibitem[Press et al.(1992)]{Pre92}
Press, W.H., Teukolsky, S.A., Vetterling, W.T., \& Flannery, B.P.
1992, Numerical Recipes in C: The Art of Scientific Computing
(Ed. 2; Cambridge: Cambridge University Press)

\bibitem[Revnivtsev et al.(2000)]{Rev00}
Revnivtsev M.~G., Trudolyubov S.~P., Borozdin K.~N., 2000, MNRAS, 312, 151

\bibitem[Rodr\'\i guez \& Mirabel(1998)]{Rod98}
Rodr\'\i guez L.~F., Mirabel I.~F., 1998, A\&A, 340, L47

\bibitem[Rodr\'\i guez \& Mirabel(1999)]{Rod99}
Rodr\'\i guez L.~F., Mirabel I.~F., 1999, ApJ, 511, 398

\bibitem[Truss \& Done(2006)]{Tru06}
Truss M., Done C., 2006, MNRAS, 368, L25

\bibitem[van Kerkwijk et al.(1996)]{van96}
van Kerkwijk M.~H., Geballe T.~R., King D.~L., van der Klis M., van
Paradijs J., 1996, A\&A, 314, 521

\end{thebibliography}

\end{document}